\documentclass[conference]{IEEEtran}

\usepackage{cite}
\usepackage{amsmath,amssymb,amsfonts}
\usepackage{algorithmic}
\usepackage{graphicx}
\usepackage{textcomp}
\usepackage{xcolor}
\usepackage{bm}
\usepackage{comment}
\usepackage{float}
\usepackage{svg}
\usepackage{color, soul}
\usepackage{balance}

\title{User Localization with HRIS and Backscatter Modulation for Next-Generation Networks}
\author{Mattia Piana and Stefano Tomasin \thanks{This work was partially supported by the European Union under the Italian
National Recovery and Resilience Plan (NRRP) of NextGenerationEU,
partnership on “Telecommunications of the Future” (PE0000001 - program
“RESTART”).} \\
\small
Dept. of Information Engineering, University of Padova, Italy \\
mattia.piana@studenti.unipd.it and stefano.tomasin@unipd.it}

\IEEEoverridecommandlockouts 
\begin{document}

\maketitle

\begin{abstract}
Hybrid reflective intelligent surfaces (HRISs) can support localization in sixth-generation (6G) networks thanks to their ability to generate narrow beams and at the same time receive and process locally the impinging signals. In this paper, we propose a novel protocol for user localization in a network with an HRIS. The protocol includes two steps. In the first step, the HRIS operates in full absorption mode and the user equipment (UE) transmits a signal that is locally processed at the HRIS to estimate the angle of arrival (AoA). In the second step, the base station transmits a downlink reference signal to the UE, and the HRIS superimposes a message by a backscatter modulation. The message contains information on the previously estimated AoA. Lastly, the UE, knowing the position of the HRIS, estimates the time of flight (ToF) from the signal of the second step and demodulates the information on the AoA to obtain an estimate of its location. Numerical results confirm the effectiveness of the proposed solution, also in comparison with the Cram\'er Rao lower bound on the estimated quantities.
\end{abstract}

\section{Introduction}

Accurate user equipment (UE) localization is a key target for next-generation cellular networks \cite{9976205} to provide new services to the UE. This target can be facilitated by new technologies that will be deployed in sixth-generation (6G) networks. In particular, reconfigurable intelligent surfaces (RISs) are seen as a transformative technology that can control the physical propagation environment in which they are embedded by passively reflecting radio waves in preferred directions \cite{wymeersch2020radio}. UE localization and sensing are also among the applications that can benefit from RIS deployments. In particular, integrated sensing and communications (ISAC) systems can benefit from RISs as shaping the wireless environment can increase the target's illumination and facilitate its detection by the UEs in radar applications \cite{liu2023integrated}.  

In this paper, we focus on the localization of the UE at the UE itself. Localization at the UE in RIS-aided scenarios was studied for base stations (BSs) and UEs with single antennas in \cite{keykhosravi2021siso}, for BSs with multiple antennas and UEs with single antennas in \cite{fascista2021ris}, and for both BS and UE with multiple antennas in~\cite{he2020large}. The localization of UE by the BS in a no line of sight (nLOS) scenario is particularly challenging due to the interaction of the RIS with the channel and the limited observation (channel matrix size) of the cascade channel through the RIS~\cite{swindlehurst2022channel}. 

While RISs are equipped with a large number of elements and can create narrow electromagnetic beams, the channel seen by the RISs (and the entailed information on UE location) remains largely unknown to both the BS and the UE, as they are equipped with much fewer antennas. To overcome this issue, a new technological solution is provided by hybrid RISs (HRISs) that, beyond reflecting, also capture part of the impinging signal for local processing~\cite{alexandropoulos2023hybrid}. Early works have started exploring the use of HRISs for localization. In \cite{ghazalian2023joint} the UE and HRIS positions are jointly estimated by the UE while in \cite{he2023joint} both BS-HRIS and HRIS-UE channels are estimated by the BS, as well as the angle of arrival at the HRIS. In both works, the HRIS actively shares with the BS its locally gathered information to further process it, via a control link. It can be argued that while sharing the local information with the BS is easy, as an infrastructure connecting the HRIS to the BS is expected to be deployed (see for example \cite{ghazalian2023joint}), sharing the local information with the UE is not straightforward.

In this paper, we propose a novel localization technique by the UE, which exploits the sensing capability of an HRIS. In our scheme, the UE transmits signals to the HRIS that operates in full absorption mode, i.e., the received signal is not reflected but fully processed locally to estimate the angle of arrival (AoA) at the HRIS. Then, after a preset time, the BS transmits a downlink signal through the HRIS that enables the UE to estimate the time of flight (ToF) from the HRIS. Moreover, in this second phase, the HRIS transmits to the UE the information on the AoA estimated in the first phase by a backscattered modulation superimposed on the signal transmitted by the BS. The UE, knowing the location of the HRIS, can then estimate its position. We denote the resulting localization protocol as localization with HRIS and backscattered signal (LHBS). LHBS exploits the control channel between the HRIS and the BS only to trigger the transmission of the downlink reference signal in the second phase. Lastly, the assumption that the UE knows the position of the HRIS is also reasonable since the HRIS is fixed and is part of the network infrastructure, thus being equivalent to that of knowing the position of BSs. 
 
The rest of this work is organized as follows: Section~\ref{sec:System Model} introduces the System model. The proposed LHBS protocol is described in Section~\ref{sec:positioning_alg}. The performance of the proposed solution is investigated in Section~\ref{sec:num_res}, where we also provide a comparison with the Cram\'er-Rao lower bound on the AoA, the range, and the position estimators. Lastly, in Section~\ref{sec:conclusions} we draw the main conclusions.

\section{System Model} \label{sec:System Model}

\begin{figure} 
    \centering
    \includegraphics[width=1\hsize]{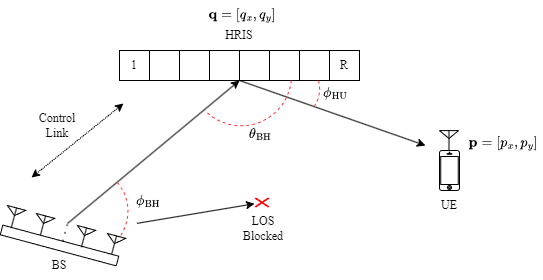}
    \caption{
    Reference scenario: an UE at position $\bm{p}$ localizes itself using an HRIS at position $\bm{q}$ and a BS in a nLOS scenario. The AoA and AoD to and from the HRIS are $\theta_{\rm BH}$ and $\phi_{\rm HU}$, respectively. The AoD from the BS is $\phi_{\rm BH}$. }
    \label{fig:schema}
\end{figure}
We consider the scenario of Fig. \ref{fig:schema} in which a UE estimates its position in a cellular network assisted by a single BS. The estimation is assisted by an HRIS controlled by the BS.

The BS is located at the origin of our reference system. The HRIS is located at position $\bm{q} = [q_x, q_y]$, which is known to the UE. The UE aims at estimating its position $\bm{p} = [p_x, p_y]$. 

The BS is equipped with a uniform linear array (ULA) with $M$ antennas, while the UE is equipped with a single antenna. The HRIS comprises $R$ evenly spaced elements organized along a line, each receiving and reflecting signals as an omnidirectional antenna while introducing a configurable phase shift. The HRIS is hybrid because it can also process impinging signals locally. The coefficient $\rho \in [ 0,1]$ represents the fraction of the signal power that is reflected, while a fraction $(1- \rho)$ of the signal power is absorbed for local processing purposes by each HRIS element. We assume that the HRIS has a direct/control link (e.g., via optical fiber) with the BS but it is not equipped with a transmitter. Consequently, the HRIS cannot directly communicate with the UE but it can only process locally the impinging signals and change its element phase shifts. We define two extreme operative modes of the HRIS, namely:
\begin{itemize}
    \item \emph{Full Absorption Mode}: in this case $\rho=0$, meaning that the whole incoming signal is absorbed by the HRIS for local processing.
    \item \emph{Fully Reflective Mode}: in this case $\rho=1$, so the whole incoming signal is reflected by the HRIS.
\end{itemize}

We consider a scenario without strict synchronization between the BS, the HRIS, and the UE. In the following, time is referred to the system of the UE, without loss of generality. We denote with $\epsilon$ the normalized (to $\lambda/(2\pi c)$) time offset between the UE and the HRIS or BS clocks, where $\lambda$ is the wavelength, and $c$ is the speed of light in air.

\subsection{Channel Model}

\paragraph*{Geometrical Properties of Signal Propagation} Considering a transmission at mmWaves, we assume that each channel is characterized by a single path. The AoA and angle of departure (AoD) to and from the HRIS are denoted as $\theta_{\rm BH}$ and $\phi_{\rm HU}$, respectively. The AoD from the BS is $\phi_{\rm BH}$. We define the ToF between the HRIS and the UE as $\tau_{\rm HU}$, consequently the distance between UE and HRIS is
\begin{equation}\label{rtau}
r_{\rm HU}= c\tau_{\rm HU}.
\end{equation}
Similarly, defining the ToF between the HRIS and the BS as $\tau_{\rm HB}$, the distance between BS and HRIS is $r_{\rm HB}= c\tau_{\rm HB}$. Let $d_{\rm tot}=r_{\rm HU} + r_{\rm HB}$ be the total distance of the downlink path, i.e., the distance that the signal sent from the BS and reflected from the HRIS travels to reach the UE. We denote as $\tau=d_{\rm tot}/c$ the ToF between the BS and the UE.  

\paragraph*{Baseband Channel Model} The free-space path loss between the UE and the HRIS is $\gamma = \left (\frac{\lambda}{4\pi r_{\rm HU}}\right )^{2}$. We define the channel coefficient from the UE to the HRIS as $\xi_{\rm HU}=\sqrt{\gamma} e^{-j \left(\epsilon +\frac{2\pi r_{\rm HU}}{\lambda}\right)} $, including the effect of path loss,   group delay between the UE and the HRIS, and clock differences. Similarly, we define $\xi_{\rm BH}$ as the channel coefficient from the BS and the HRIS similarly defined (using $r_{\rm hb}$). Lastly, we define the steering vector operator at the receiver as
\begin{equation}
\bm{a}_A(\theta) = [1,e^{-j\kappa\sin{\theta}}, \ldots ,e^{-j\kappa\sin{\theta}(A-1)}]^T,
\end{equation}
where $\kappa=2\pi d /\lambda$, $d$ is the spacing between two adjacent antennas (or HRIS elements), $A \in \{M,R\}$  is the number of antennas (or elements), and $\theta$ represents the AoA. The baseband single-path channel matrix from the BS to the HRIS is therefore
\begin{equation}
\bm{H}_{\rm BH} = \frac{1}{\sqrt{M}}\xi_{\rm BH} \, \bm{a}_R(\theta_{\rm BH})\bm{a}_M^H(\phi_{\rm BH}) \in \mathbb{C}^{R \times M}.
\end{equation}
The column-vector channel from the UE to the HRIS is
\begin{equation} \label{eq:channel_ue_hris}
\bm{h}_{\rm UH} = \xi_{\rm UH} \, \bm{a}_R(\phi_{\rm HU}) \in \mathbb{C}^{R \times 1}.
\end{equation}
The BS-HRIS channel $\bm{H}_{\rm BH}$ is assumed to be known both at the BS and at the HRIS as it is fully determined by their positions. Let $\bm{b}_{\rm BS} \in \mathbb{C}^{M}$ be the beamforming vector at the BS. 

\paragraph*{HRIS Model} We characterize the effect of the HRIS by the diagonal matrix $\bm{\Omega} \in \mathbb{C}^{R \times R}$, where each element introduces the following gain and phase shift
\begin{equation}
\bm{\Omega}_{ii}=\sqrt{\rho} e^{j\alpha_i}, \quad \alpha_i \in [0,2\pi),\; i = 0, \ldots , R-1.
\end{equation}
We assume that the line of sight (LOS) between the UE and the BS is blocked, so the communication is assisted by an HRIS.




\section{Localization With HRIS \\ and Backscattered Signal} \label{sec:positioning_alg}

In this section, we describe the localization with HRIS and backscattered signal (LHBS) at the UE. Our approach is based on the estimation of two quantities: the AoD at the HRIS $\phi_{\rm HU}$ and the ToF $\tau_{\rm HU}$. From this information, the UE will estimate its position.
Our algorithm is divided into three phases, which are first briefly outlined and then explained in detail:
\begin{itemize}
   \item Phase 1- \emph{AoA estimation at the HRIS}: the single-antenna UE transmits a pilot signal. The HRIS, in full absorption mode, detects the signal coming from the UE and estimates its AoA. When the HRIS detects the UE pilot signal, it notifies the BS of the reception via the control link.
   \item Phase 2- \emph{Acknowledgment}: After a pre-agreed time, the BS sequentially transmits two pilot signals that are fully reflected through the HRIS to the UE. In this phase, the HRIS sets its phases according to the estimated AoA in the first phase to steer the BS signal towards the UE.  Moreover, the reflected signals are modulated by the HRIS (in backscattering) to convey the information regarding  $\phi_{\rm HU}$. 
   \item Phase 3- \emph{UE position estimation}: when the UE receives the reflected signals, it demodulates the AoD $\phi_{\rm HU}$ and estimates the ToF $\tau_{\rm HU}$. With this information, it finally estimates its position.
\end{itemize}

\begin{figure} 
    \centering
    \includegraphics[width=1\hsize]{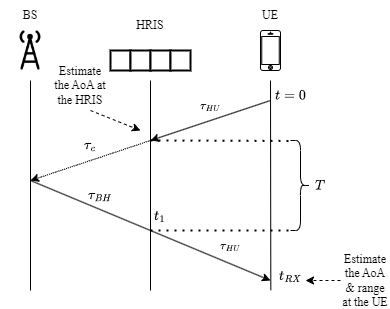}
    \caption{Timing scheme of the LHBS protocol: $\tau_c$ is the control-link propagation time, $\tau_{\rm HU}$ is the ToF between UE and HRIS, and $\tau_{\rm BH}$ is the ToF between BS and HRIS. Processing times are ignored.}
    \label{fig:timing_schema}
\end{figure}

The timing scheme of the LHBS protocol is summarized in Fig.~\ref{fig:timing_schema}. Note that the processing times are ignored but their inclusion in our model is straightforward.

\subsection{AoA Estimation at the HRIS} \label{sec:phase_1}

The LHBS procedure starts (at time $t=0$, UE time) with the UE transmitting the reference sequence. This sequence is used at the HRIS to detect the impinging signal by correlating it with a local replica and to estimate the AoA from the UE. We assume here that the HRIS operates in full absorption mode. 
When the HRIS detects the signal, it notifies the BS of the reception via the control link, the latter will use this information in the next phase. 

In detail, the UE transmits a reference signal of $N$ symbols $\bm{x} = [x(0), \ldots,x(N-1)]^T$, using a unit-power square root raised cosine (SRRC) pulse $q(t)$, with symbol period $T_c$. Without loss of generality, we assume $N$ to be even. We use a constant-amplitude zero-autocorrelation (CAZAC) sequence \cite{albanese2021papir}, where
\begin{equation}
x_n = e^{ j\pi \frac{n(n + 1)}{N}}, \quad n=0, \ldots ,N-1.
\end{equation}
This specific sequence, having good autocorrelation properties, allows for precisely detecting the time of arrival of the signal at the HRIS.

At the HRIS, the received signal is filtered with a filter with impulse response $q(-t)$ and sampled at period $T_c$ (with a sample falling in correspondence with the peak of the convolution between $q(t)$ and $q(-t)$), providing 
\begin{equation} \label{eq:sig_at_HRIS}
    \bm{y}_{\rm H} (n+k') 
    =\bm{h}_{\rm UH} x_n+\bm{w}_{\rm H}(n+k') \in \mathbb{C}^{R\times 1},        
\end{equation}
where entries of noise vector $\bm{w}_{\rm H}(n)$ are independent and identically distributed as  $\mathcal{CN}(0,\,\sigma^{2})$, and $k'$ is a suitable delay due to both propagation and sampling delays.
We notice that \eqref{eq:sig_at_HRIS} is a typical signal model for the multiple signal classification (MUSIC) algorithm \cite[Sec. 8.6.3]{hayes1996statistical} that provides the estimate $\hat{\phi}_{\rm HU}$ of the AoA. Hence, the HRIS estimates the AoA.

\subsection{Acknowledgment} \label{sec:phase_2}

In the second phase of LHBS, the HRIS modulates the information on the estimated AoA by backscattering on a pilot signal transmitted by the BS. The modulated signal also enables the UE to estimate the ToF from the HRIS to the UE. In detail, the backscatter transmission is obtained by the modulation of the incoming signal with two pilot sequences by the HRIS, which will enable the demodulation of the backscattered information by differential demodulation. This solution avoids that the UE needs to know in advance the channel from the BS to the UE.

In our protocol, also the BS will transmit a stream of symbols $x_n \in \mathbb{C}$ using an SRRC pulse $q(t)$ with symbol period $T_c$. In this phase, the HRIS works in fully reflective mode. The BS aligns its beamformers to the HRIS direction by setting 
\begin{equation} \label{eq:beamformers}
\bm{b}_{\rm BS}=\frac{1}{\sqrt{M}}\bm{a}_M(\phi_{\rm BH}).
\end{equation}

\paragraph*{First Transmission} In transmitting the first acknowledgment signal, the HRIS will convey the information on the AoA by a backscatter modulation. In particular, in this phase, the HRIS will use this information to set its phase shifters as
\begin{equation} \label{eq:general_HRIS_phase}
    \bm{\Omega}_{ii} = e^{jk[\sin{\hat{\phi}_{\rm HU}}-\sin{\theta_{\rm BH}}]i}e^{j\hat{\phi}_{\rm HU}}.
\end{equation}
The timing of the first transmission is such the HRIS receives the first sample at time $t_1 = \kappa_1 T_c$, so time $T$ is elapsed from the reception at the HRIS of the UE signal of phase 1. Using \eqref{eq:beamformers} and \eqref{eq:general_HRIS_phase} we obtain  (see Fig. \ref{fig:timing_schema})
\begin{equation} \label{eq:cont_time_baseband}
\begin{split}
    r_{\rm UE}(t)= & R\xi_{\rm BH}\xi_{\rm UH}e^{j\hat{\phi}_{\rm HU}} \times \\
    & \times \sum_{n=0}^{N-1} x_nq(t-nT_c-2\tau_{\rm HU}-T)+ \\
    & +w'(t),
\end{split}
\end{equation}
where $w'(t)$ is the AWGN and both AoA and ToF appear. This model will be used in Section~\ref{sec:cramer} to assess the AoA and range estimator performance. Then the UE filters the received signal by $q(-t)$ and upsamples the filter output with over-sampling period  $T_s=\frac{T_c}{u_{\rm f}}$, where $u_{\rm f}$ is the oversampling factor. Let us define
\begin{equation}
   s_i=  R\xi_{\rm BH}\xi_{\rm UH}e^{j\hat{\phi}_{\rm HU}} \sum_{n=0}^{N-1} x_n \tilde{q}(iT_s-nT_c-2\tau_{\rm HU} - T-\tau_1), 
\end{equation}
for $i=0,\ldots, Nu_{\rm f}-1$, where $\tilde{q}(t)$ is the convolution between $q(t)$ and $q(-t)$, and $\tau_1$ is a suitable delay so that the maximum of the convolution is one of the samples of $s_i$ and all non-zero terms fall in the considered window of $i$. The oversampled signal model is
\begin{equation} \label{eq:upsampled_time_baseband}
    \bm{r}= \bm{s}+\bm{w},
\end{equation}
where entries of $\bm{r}$ are $r_i=  s_i+w(iT_s)$ and $w(iT_s) \sim \mathcal{CN}(0,\,\sigma^{2}_{\rm u})$ for $i=0,...,Nu_{\rm f}-1$. Note that, $\sigma^{2}_{\rm u} = u_{\rm f} \sigma^{2}$ as up sampling the received signal also increases the noise variance. Recalling \eqref{rtau}, we notice that $\bm{r}\sim \mathcal{CN}(\bm{s}(\hat{\phi}_{\rm HU},r_{\rm HU}),\,\sigma^{2}_{\rm u} \bm{I}_{Nu_{\rm f}})$, i.e., the range and AoA information is contained in the mean of a Gaussian vector.


Note that we could obtain the same goal via digital modulation of the HRIS elements. As an example, in \cite{yan2020passive} a passive (non-hybrid) RIS is exploited and the state of its elements (on/off) is used to convey its local information via a spatial modulation scheme. 

The estimation of the AoA at the HRIS in phase 1 is crucial, as it will affect the modulation that HRIS will apply in phase~2. Since the path loss of the signal received by the HRIS when estimating the UE AoA is much lower than the one that affects the BS-HRIS-UE path, we can take the estimation at the HRIS as perfect. In other words, due to the large number of HRIS elements and the mmWave frequencies, if we want our algorithm to work then the estimation error at the HRIS can be made much smaller with respect to the end-to-end one. 
Hence, the bottleneck of the system is the estimation at the UE of the AoA with the signal coming from the BS and reflected by the HRIS.

\paragraph*{Second Transmission} In the transmission of the second acknowledgment signal, the HRIS optimally aligns the channel between BS and UE and creates a virtual LOS between the two. To do that, the HRIS sets its phase shifters as
    \begin{equation} \label{eq:HRIS_phase_aligned}
    \bm{\Omega}_{ii} = e^{jk[\sin{\hat{\phi}_{\rm HU}}-\sin{\theta_{\rm BH}}]i}.
\end{equation}

In this case, no additional modulation is added by the HRIS. The second transmission starts immediately after the first and the received discrete-time signal at the UE (after filtering and sampling) can be written similarly to \eqref{eq:upsampled_time_baseband} as
\begin{equation} \label{eq:upsampled_time_baseband2}
    \bm{r}'= \bm{s}'+\bm{w}',
\end{equation}
where $\bm{w}'$ is the AWGN vector and entries of $\bm{s}'$ are 
\begin{equation}
   s_i'=  R\xi_{\rm BH}\xi_{\rm UH} \sum_{n=0}^{N-1} x_n \tilde{q}(iT_s-nT_c-2\tau_{\rm HU} - T-\tau_1-NT_c), 
\end{equation}
for $i=0,\ldots, Nu_{\rm f}-1$.


The UE demodulates $\hat{\phi}_{\rm HU}$ by aligning in time the two signals, and then combining them as follows
\begin{equation} \label{eq:differential_decoding}
\begin{split}
    z &= \frac{1}{Nu_{\rm f}} \sum_{n=0}^{Nu_{\rm f}-1} r_{n}^* r'_n \\
       & \simeq||\xi_{\rm BH}\xi_{\rm UH}||^2 R^2  e^{-j\hat{\phi}_{\rm HU}}.
\end{split}
\end{equation}
The approximation in (\ref{eq:differential_decoding}) arises from the large number theorem as we substituted the average with the expectation operator when $N$ is sufficiently large. The UE can now obtain the analog demodulated AoA at the HRIS as
\begin{equation} \label{eq:theta_estim_bs}
    \dot{\phi}_{\rm HU} =- \angle{z}.
\end{equation}
Note that in both transmissions we use a CAZAC sequence of length $N$. In principle, we could use a general pilot sequence not necessarily known at the UE to estimate the AoA, as (\ref{eq:differential_decoding}) holds when we use the same sequence in both transmissions. Still, we will use these two sequences also in Section~\ref{sec:phase_3} for the range estimation.

\subsection{UE Position Estimation} \label{sec:phase_3}

In the third phase of LHBS, the UE computes its position. In particular, the UE knows the time of its transmission (time zero) and also the fixed time $T$. Consequently, the UE can extract the ToF $\tau_{\rm HU}$ from both the received signals of the first and second transmission by the BS, and then take the average. Note the estimate of the dalay cannot be done on the two sequences together, since they have an unknown phase shift, due to backscatter modulation.

In detail, the UE correlates the over-sampled signal with a local replica of the over-sampled transmitted sequence. The relation between the transmitting and arrival times at the UE is (e.g., for the first sequence)
\begin{equation} \label{eq:arrival_time}
    t_{ \rm RX} = 2\tau_{\rm HU}+T,
 \end{equation}
 and so a first estimate of the ToF at the UE is
  \begin{equation} \label{eq:tof_estimation}
     \tilde{\tau}_{\rm HU} = \frac{t_{ \rm RX}-T}{2}.
 \end{equation}
Similarly, we obtain the second estimation $\tilde{\tau}'_{\rm HU}$ from the second transmission. Lastly, the estimation of the range at the UE is obtained by computing 
\begin{equation}
    \dot{r}_{\rm HU}=c \left (\frac{\tilde{\tau}_{\rm HU}+\tilde{\tau}'_{\rm HU}}{2} \right ).
\end{equation}
Note that estimating the range in this way relies only on the absolute times at the UE and is not affected by clock misalignment between UE and BS.

With the AoA and range estimations $\dot{\phi}_{\rm HU}$ and $\dot{r}_{\rm HU}$, and the known HRIS position $\bm{q}$, the UE estimates its position $\dot{\bm{p}}$. In particular, defining $\bm{\theta}= [\phi_{\rm HU},r_{\rm HU}]^T$, we have 
\begin{equation} \label{eq:final_position}
\left[\begin{matrix}
\dot{p}_x\\
\dot{p}_y
\end{matrix}\right]=  \bm{g}(\dot{\bm{\theta}}) = \bm{q} + \left[
\begin{matrix}
\, \, \, \,\dot{r}_{\rm HU}\cos{\dot{\phi}_{\rm HU}} \\
-\dot{r}_{\rm HU}\sin{\dot{\phi}_{\rm HU}}
\end{matrix}\right].
\end{equation}
 
\subsection{Cram\'er-Rao Lower Bounds} \label{sec:cramer}

\begin{figure}
     \centering
         \includegraphics[width=1\hsize]{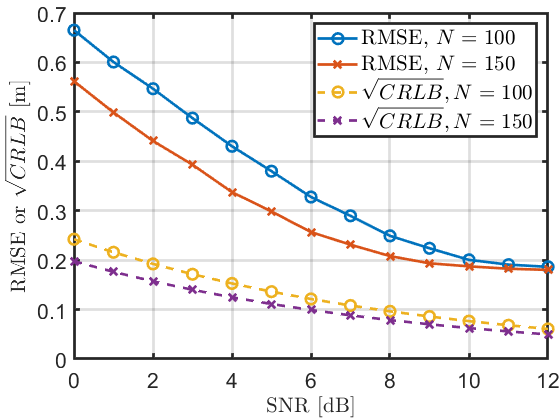}
         \caption{RMSE and $\sqrt{CRLB}$ of the ranging estimate vs the SNR for a sequence length $N=100$ or 150.}
         \label{fig:raging_error}
\end{figure}

\begin{figure}
    \centering
    \includegraphics[width=1\hsize]{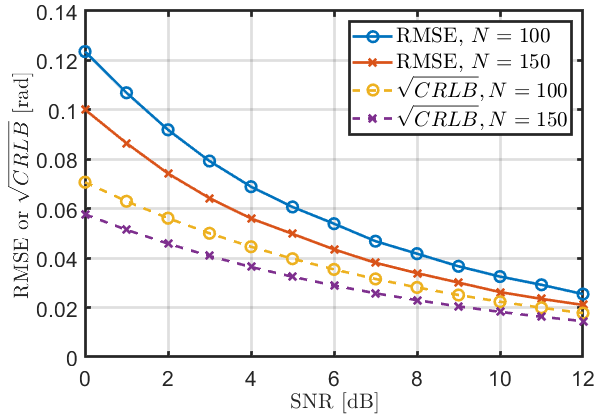}
    \caption{RMSE and $\sqrt{CRLB}$ of the AoA estimate vs the SNR for a sequence length $N=100$ or 150.}
    \label{fig:ue_estimation_error}
\end{figure}

To assess the performance of the LHBS algorithm, we compared the MSE of the range $r_{\rm HU}$ and the demodulated AoA $ \dot{\phi}_{\rm HU}$ with their Cram\'er Rao lower bounds (CRLB) that give a lower bound on the error variance of all the possible unbiased estimators of the parameters. Due to space limitations, the algebraic operations to compute the quantities in the following can be found in the Appendix.


\paragraph*{CRLB of the Received AoA} We assume that the estimate of the AoA at the HRIS is without error, while we take into account the error introduced in the transmission from the BS to the UE through the HRIS. From \cite[Sec. 15.7]{kay1993fundamentals} we compute

\begin{equation}     \label{eq:cramer_dimostrazione} 
\begin{split}
       \frac{ \partial ^2 \,\ln\, p(\bm{r}|\hat{\phi}_{\rm HU},r_{\rm HU})}{\partial ^2\hat{\phi}_{\rm HU}}
   &=  \frac{2}{\sigma^{2}_{\rm u}}  \frac{\partial}{\partial \hat{\phi}_{\rm HU}} \Re \left[(\bm{r}-\bm{s}(\hat{\phi}_{\rm HU},r_{\rm HU}))^H \times \right. 
   \\ & \times \left.
 \frac{ \partial \,\bm{s}(\hat{\phi}_{\rm HU},r_{\rm HU})}{\partial \hat{\phi}_{\rm HU}}
   \right ]  = \frac{2}{\sigma^{2}_{\rm u}} \Re [ -\bm{r}^H\bm{s} ].
\end{split}
\end{equation}
The CRLB on the AoA is then
\begin{equation} \label{eq:cramer_1}
\begin{split}
CRLB_{\phi_{\rm HU}} &=-\left[\mathbb{E}\left (\frac{ \partial ^2 \,\ln\, p(\bm{r}|\hat{\phi}_{\rm HU},r_{\rm HU})}{\partial ^2\hat{\phi}_{\rm HU}} \right )\right]^{-1}   = \frac{\sigma^{2}_{\rm u}}{2||\bm{s}||^2}.
\end{split}
\end{equation}

\paragraph*{CRLB of the Ranging Estimation at the UE} We also compute the CRLB of the estimate of the ranging using the results of  \cite[Sec. 2.4.6]{mengali2013synchronization} that provide
\begin{equation} \label{eq:cramer_2}
CRLB_{r_{\rm HU}}= \frac{c^2}{8\cdot 2 B_2^2\frac{E_s}{N_0}},
\end{equation}
where $E_s$ is the total energy used for the correlation, $B_2$ is the root mean square bandwidth, and $N_0 = \sigma^2_{\rm u}T_s$. The coefficient 8 at the denominator of \eqref{eq:cramer_2} is due to our use of two sequences to perform the estimation of the ToF and both of them estimate $2\tau_{\rm HU}$. 
\paragraph*{CRLB of the Position Estimation} To obtain the CRLB of the position estimation we first compute the Fisher information matrix (FIM) on the AoA and range.  In detail, the FIM is 
\begin{equation}
    \bm{I} = \begin{bmatrix} 
CRLB_{\phi_{\rm HU}}^{-1} & \alpha  \\
\alpha & CRLB_{r_{\rm HU}}^{-1}\end{bmatrix}.
\end{equation}
where
\begin{equation}
\begin{split}      
    \alpha &= -\mathbb{E}\left (\frac{ \partial ^2 \,\ln\, p(\bm{r}|\hat{\phi}_{\rm HU},r_{\rm HU})}{\partial \hat{\phi}_{\rm HU} \partial r_{\rm HU}}  \right ) \\ 
   &= -\frac{2}{\sigma^{2}_{\rm u}}\Im \left [\bm{s}^H \frac{\partial \bm{s}}{\partial r_{\rm HU}}  \right ].
\end{split}
\end{equation}
From \cite[Sec. 3.8] {kay1993fundamentals} we obtain the CRLB on the position estimation at the UE as
\begin{equation}\label{eq:mse_final_position}
    CRLB_{\rm pos}={\rm tr}\left [ \frac{\partial \bm{g}(\bm{\theta}) }{\partial \bm{\theta}} \bm{I}^{-1}\frac{\partial \bm{g}(\bm{\theta})^T }{\partial \bm{\theta} } \right] .
\end{equation}

\begin{figure}
         \centering
         \includegraphics[width=1\hsize]{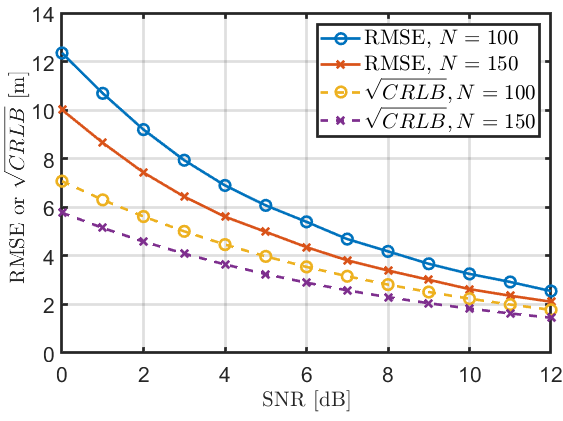}
         \caption{RMSE or $\sqrt{CRLB}$ of the position estimation vs SNR for a sequence length of $N=100$ or 150.}
         \label{fig:final_position_error}
\end{figure}
\section{Numerical Results} \label{sec:num_res}

In this section, we assess the performance of the proposed LHBS protocol and compare it with the obtained CRLBs. The transmit power is $\mathbb{E}(||x_n||^2)=1$, the carrier frequency is $f_c= 25$ GHz, the distance HRIS-UE is $r_{\rm HU}=100 $ m, the distance BS-HRIS is $r_{\rm BH}=100$ m, the sequence length is $N=100$ or $150$, the signal bandwidth is $B=20$ MHz, the roll-off factor is $\beta=0.8$, and the up sampling factor is $u_{\rm f}=10$.  

We define the  signal-to-noise ratio (SNR) on the received signal as 
\begin{equation} \label{eq:SNR_def}
    \mathrm{SNR} =  \frac{||\xi_{\rm BH}\xi_{\rm UH}||^2  R^2}{\sigma^2}. 
\end{equation}


Fig.~\ref{fig:raging_error} shows the root mean square error (RMSE) of the ranging $r_{\rm HU}$ as a function of the SNR. We also plot the corresponding square root of CRLB \eqref{eq:cramer_2}. We observe that the RMSE of the ranging error is in the range of tens of centimeters and for high SNR values the performance of the estimator gets close to the CRLB. Also, we observe that increasing the sequence length ($N$) provides a performance improvement. We also observe that the improvement saturates at high SNR: this is due to the time quantization. Upsampling even more the received signal would lower the saturation level.

Fig.~\ref{fig:ue_estimation_error} shows the RMSE of the AoA estimated at the HRIS as a function of the SNR, again in a comparison with the square root of the CRLB \eqref{eq:cramer_1}. We note that in this case, the estimator achieves a performance that is closer to the CRLB.

Lastly, Fig.~\ref{fig:final_position_error} shows the RMSE of the estimated UE position as a function of the SNR. We observe that in the considered scenario we can achieve a positioning precision in the range of meters.
\balance
\section{Conclusions} \label{sec:conclusions}
In this paper, we presented a novel protocol for HRIS-aided user localization. We exploited the sensing property available at the HRIS to locally process the incoming signal from the UE and estimate the AoA of the signal. The UE then computes its position from two signals transmitted from the BS and reflected by the HRIS. They encode by a differential backscatter modulation the AoA information. We proved the effectiveness of our approach via simulation and compared the performances with the Cram\'er Rao lower bounds.

\bibliographystyle{IEEEtran}
\bibliography{HRIS_bib}

\onecolumn
\section{Appendix}
\subsection{CRLB of the Received AoA}
In this section, the CRLB of the Received AoA will be derived.
\begin{equation}     \label{eq:cramer_dimostrazione_appendix} 
\begin{split}
       \frac{ \partial ^2 \,\ln\, p(\bm{r}|\hat{\phi}_{\rm HU},r_{\rm HU})}{\partial ^2\hat{\phi}_{\rm HU}}
   &=  \frac{2}{\sigma^{2}_{\rm u}}  \frac{\partial}{\partial \hat{\phi}_{\rm HU}} \Re \left[(\bm{r}-\bm{s}(\hat{\phi}_{\rm HU},r_{\rm HU}))^H \frac{ \partial \,\bm{s}(\hat{\phi}_{\rm HU},r_{\rm HU})}{\partial \hat{\phi}_{\rm HU}} \right ] \\
   &=  \frac{2}{\sigma^{2}_{\rm u}}  \frac{\partial}{\partial \hat{\phi}_{\rm HU}} \Re \left[(\bm{r}^H-\bm{s}^H(\hat{\phi}_{\rm HU},r_{\rm HU}))
j \bm{s}(\hat{\phi}_{\rm HU},r_{\rm HU})
   \right ] \\
   &=  \frac{2}{\sigma^{2}_{\rm u}}  \frac{\partial}{\partial \hat{\phi}_{\rm HU}} \Re \left[j \, \bm{r}^H\bm{s}(\hat{\phi}_{\rm HU},r_{\rm HU})
   \right ] \\
   &= \frac{2}{\sigma^{2}_{\rm u}} \Re [ -\bm{r}^H\bm{s}(\hat{\phi}_{\rm HU},r_{\rm HU}) ].
\end{split}
\end{equation}
If we take the expectation of \eqref{eq:cramer_dimostrazione_appendix} we simply substitute  $\bm{r}$ with $\bm{s}(\hat{\phi}_{\rm HU},r_{\rm HU})$, and from that the CRLB immediately follows. 
\subsection{Off-diagonal coefficient}
In this section, the off-diagonal coefficient
\begin{equation}
    \alpha = -\mathbb{E}\left (\frac{ \partial ^2 \,\ln\, p(\bm{r}|\hat{\phi}_{\rm HU},r_{\rm HU})}{\partial \hat{\phi}_{\rm HU} \partial r_{\rm HU}}  \right ) 
\end{equation}
is derived. We start by computing
\begin{equation} \label{eq:incubo_dei_matematici}
    \frac{ \partial ^2 \,\ln\, p(\bm{r}|\hat{\phi}_{\rm HU},r_{\rm HU})}{\partial \hat{\phi}_{\rm HU} \partial r_{\rm HU}} 
    = \frac{2}{\sigma^{2}_{\rm u}}  \frac{\partial}{\partial \hat{\phi}_{\rm HU}} \Re \left[(\bm{r}-\bm{s}(\hat{\phi}_{\rm HU},r_{\rm HU}))^H
 \frac{ \partial \,\bm{s}(\hat{\phi}_{\rm HU},r_{\rm HU})}{\partial r_{\rm HU}}
   \right ].
\end{equation}
Knowing that $\partial r_{\rm HU}=c \, \partial \tau_{\rm HU}$ , we rewrite \eqref{eq:incubo_dei_matematici} as
\begin{equation} \label{eq:uno}
    \frac{ \partial ^2 \,\ln\, p(\bm{r}|\hat{\phi}_{\rm HU},r_{\rm HU})}{\partial \hat{\phi}_{\rm HU} \partial r_{\rm HU}} 
    = \frac{2}{\sigma^{2}_{\rm u}}  \frac{\partial}{\partial \hat{\phi}_{\rm HU}} \Re \left[(\bm{r}-\bm{s}(\hat{\phi}_{\rm HU},r_{\rm HU}))^H
 \frac{ \partial \,\bm{s}(\hat{\phi}_{\rm HU},r_{\rm HU})}{c \, \partial \tau_{\rm HU}}
   \right ].
\end{equation}
We compute $\check{\bm{s}}$ with entries
\begin{equation}
    \check{s}_i=\left [\frac{ \partial \,\bm{s}(\hat{\phi}_{\rm HU},r_{\rm HU})}{\, \partial \tau_{\rm HU}} \right ]_i = -2 R\xi_{\rm BH}\xi_{\rm UH}e^{j\hat{\phi}_{\rm HU}} \sum_{n=0}^{N-1} x_n \tilde{q}'(iT_s-nT_c-2\tau_{\rm HU} - T-\tau_1),
\end{equation}
where $\tilde{q}'(\cdot)$ is the raised cosine derivative.
We notice that $\bm{s}(\hat{\phi}_{\rm HU},r_{\rm HU})^H\frac{ \partial \,\bm{s}(\hat{\phi}_{\rm HU},r_{\rm HU})}{\, \partial \tau_{\rm HU}}$ does not depend on $\hat{\phi}_{\rm HU}$, consequently its derivative with respect to $\hat{\phi}_{\rm HU}$ is zero. Thus, we rewrite \eqref{eq:uno} as
\begin{equation} \label{eq:due}
\begin{split}
    \frac{ \partial ^2 \,\ln\, p(\bm{r}|\hat{\phi}_{\rm HU},r_{\rm HU})}{\partial \hat{\phi}_{\rm HU} \partial r_{\rm HU}} 
    &= \frac{2}{\sigma^{2}_{\rm u}}  \frac{\partial}{\partial \hat{\phi}_{\rm HU}} \Re \left[\bm{r}^H
 \frac{ \partial \,\bm{s}(\hat{\phi}_{\rm HU},r_{\rm HU})}{c \, \partial \tau_{\rm HU}}
   \right ]\\
   &= \frac{2}{\sigma^{2}_{\rm u}}\Re \left[ j \, \bm{r}^H
 \frac{ \partial \,\bm{s}(\hat{\phi}_{\rm HU},r_{\rm HU})}{c \, \partial \tau_{\rm HU}}
   \right ].
\end{split}
\end{equation}
Finally, we find
\begin{equation}
\begin{split}      
    \alpha &= -\mathbb{E}\left (\frac{ \partial ^2 \,\ln\, p(\bm{r}|\hat{\phi}_{\rm HU},r_{\rm HU})}{\partial \hat{\phi}_{\rm HU} \partial r_{\rm HU}}  \right ) \\ 
   &=-\frac{2}{\sigma^{2}_{\rm u}}\Im \left [\bm{s}^H \check{\bm{s}}\right ]  = -\frac{2}{\sigma^{2}_{\rm u}}\Im \left [\bm{s}^H \frac{\partial \bm{s}(\hat{\phi}_{\rm HU},r_{\rm HU})}{\partial r_{\rm HU}}  \right ].
\end{split}
\end{equation}

\end{document}